# Low-Temperature Stabilization of δ-NbN Superconducting Thin Films through Energy-Selective Ion Beam Sputtering

Yifan Liu[1,2], Junwoo Lee[3], Mingyu Xu[4], Weiwei Xie[4*], Robert J. Cava[5*], Qi Hua Fan[1,3*]

1. Department of Electrical and Computer Engineering, Michigan State University, East Lansing, MI 48824, USA
2. Institute for Quantitative Health Science and Engineering, Michigan State University, East Lansing, MI 48824, USA
3. Department of Chemical Engineering and Materials Science, Michigan State University, East Lansing, MI 48824, USA
4. Department of Chemistry, Michigan State University, East Lansing, MI 48824, USA
5. Department of Chemistry, Princeton University, Princeton, NJ 08540, USA

* Robert J. Cava, Qi Hua Fan, Weiwei Xie

**Email:** rcava@princeton.edu; qfan@egr.msu.edu; xieweiwe@msu.edu

**Author Contributions:** Y.L. and J.L. contributed equally to this work. R.J.C. and Q.H.F. designed research; Y.L. and J.L. performed materials synthesis and characterization; M. X. and W.X. conducted superconductivity measurements. Y.L., W.X., R.J.C. and Q.H.F. wrote the paper. R.J.C. and W.X. supervised the project. All authors reviewed the paper.

**Competing Interest Statement:** The authors declare no competing interests.

**Keywords:** Quantum information science; Superconducting thin films; Niobium nitride; Ion beam–assisted deposition; Nonequilibrium materials synthesis.

***Abstract***

Achieving high-quality superconducting thin films at low temperatures is a central challenge for scalable quantum device integration and Complementary Metal-Oxide-Semiconductor (CMOS)-compatible fabrication. Here, we demonstrate a room-temperature magnetron sputtering approach for synthesizing niobium nitride (NbN) thin films using a novel energy-selective ion source that enables independent control of ion energy and ion flux. This capability provides a powerful route to precisely tailor surface kinetics and crystallization pathways, overcoming longstanding limitations in stabilizing high-quality NbN at reduced temperatures. The resulting films were characterized by superconducting transport measurements, X-ray diffraction, X-ray photoelectron spectroscopy, and transmission electron microscopy. Compared with conventional reactive sputtering, ion beam-assisted growth yields a substantial narrowing of the transition width at the superconducting temperature, indicating improved phase purity and electronic homogeneity. Furthermore, this work introduces energy-selective ion control as a general, non-equilibrium route to stabilize metastable superconducting phases at low temperatures, offering a new design paradigm for thin-film quantum materials.

## Main Text

### Introduction

Quantum information science, QIS, is driving a new generation of technologies that exploit quantum coherence and entanglement for computation, communication, and sensing. Among the diverse material platforms under development, superconducting thin films occupy a central position because they enable low-loss microwave circuits, superconducting qubits, kinetic inductance detectors, superconducting nanowire single-photon detectors (SNSPDs), and cryogenic interconnects(1–10). As these technologies transition from laboratory demonstrations toward scalable architectures, there is an increasing demand for superconducting films that combine excellent electronic performance with manufacturing processes compatible with semiconductor fabrication(4, 7, 11, 12). Achieving this goal requires new synthesis strategies capable of precisely controlling phase formation and microstructure while remaining compatible with low-temperature processing and large-scale integration.

One of the major challenges in thin-film synthesis is controlling energy transfer during growth. Film crystallization, defect formation, and phase selection are governed by the kinetic energy delivered to surface atoms, particularly under nonequilibrium deposition conditions. Ion beam-assisted deposition provides an attractive route for engineering these processes by supplying energetic ions that enhance adatom mobility and promote crystallization well below thermodynamic equilibrium temperatures(13, 14). However, conventional ion sources typically produce broad ion energy distributions, making it difficult to independently tune ion energy and ion flux. Because these two parameters govern distinct aspects of thin-film growth, their coupling limits precise control over phase evolution and ultimately constrains the attainable material quality. An ion source capable of delivering a narrow and energy-selective ion distribution therefore offers a fundamentally new approach to engineering nonequilibrium growth pathways.

Niobium nitride (NbN) represents an ideal model system to demonstrate this concept. NbN possesses a relatively high superconducting critical temperature ($T_c$), a large upper critical magnetic field, and fast superconducting response, making it one of the most widely used materials for superconducting quantum devices, particularly SNSPDs and microwave superconducting electronics(6, 15–20). Despite decades of research, obtaining the superconducting FCC rocksalt like $\delta$-phase NbN remains challenging because conventional deposition methods—including chemical vapor deposition, thermal nitridation, and reactive magnetron sputtering—often produce mixtures of competing Nb-N phases(21–26). Stabilization of $\delta$-NbN typically requires deposition temperatures exceeding 400 °C and frequently relies on epitaxial substrates such as MgO or AlN (27–30). These requirements are incompatible with back-end Complementary Metal-Oxide-Semiconductor (CMOS) processing and limit direct integration with silicon-based quantum circuits(7, 11, 12). Despite the fact that in equilibrium the phase at a 1 to 1 ratio has a different symmetry, superconducting FCC NbN thin films have been made at or near room temperature using carefully optimized reactive sputtering conditions(31–33).

Previous studies have shown that ion beam–assisted deposition can promote room-temperature formation of $\delta$-NbN by enhancing surface diffusion and crystallization (13). Nevertheless, the broad ion energy distributions characteristic of conventional ion sources restrict precise control of the energy transferred to the growing film, limiting optimization of superconducting properties (13, 14).

Here, we introduce an energy-selective broad-beam ion source integrated with magnetron sputtering that enables independent control of ion energy and ion flux during NbN growth. The narrow ion energy distribution provides unprecedented control over the nonequilibrium crystallization process, allowing the superconducting FCC NbN to be stabilized at room temperature and at substantially lower nitrogen concentrations than required by conventional sputtering. By combining structural characterization, X-ray photoelectron spectroscopy, transmission electron microscopy, and superconducting transport measurements, we demonstrate that precise ion-energy control simultaneously improves nitridation, crystallinity, and superconducting performance, yielding NbN thin films with Tc up to about 14.8 K, narrow transition widths, and high upper critical fields. Beyond NbN, these results establish energy-selective ion engineering as a general strategy for synthesizing high-quality quantum materials under nonequilibrium conditions and provide a scalable pathway toward CMOS-compatible superconducting technologies for quantum information science(5, 7, 8, 11).

## Results and Discussion

**Energy-selective ion beam-assisted synthesis of NbN thin films:** To enable precise control over thin-film growth under non-equilibrium conditions, we employed an energy-selective ion beam source integrated with magnetron sputtering. **FIG. 1*a*** compares the ion energy distributions of a conventional anode-layer ion source and the energy-selective ion beam source, both operated at 120 eV. The energy-selective source exhibits a significantly narrower energy distribution, with a full width at half maximum (FWHM) approximately 4% of that of the conventional source, enabling substantially improved control over energy delivery to the growing film. Such control is critical for tuning surface diffusion, densification, and phase formation during deposition. NbN thin films were synthesized via ion-beam-assisted magnetron sputtering (IBAMS) under varying nitrogen concentrations and ion source conditions. For direct comparison, films were deposited both with and without ion beam assistance under otherwise identical conditions.

Under identical $N_2$/Ar ratios, films grown with ion beam assistance exhibit a higher degree of nitridation compared to those deposited by conventional sputtering. For example, at an 8% $N_2$ ratio, films deposited without ion assistance remain in a metallic Nb-dominated phase, whereas those grown with ion assistance already exhibit signatures of superconducting δ-phase FCC NbN. This behavior highlights a key advantage of the energy-selective ion beam approach: it effectively lowers the nitrogen concentration threshold required for δ-phase formation. Compared to previous ion-assisted methods that rely on broad energy distributions, the narrow energy control in this work enables more efficient energy transfer and selective phase stabilization, providing a more robust and tunable pathway for low-temperature NbN synthesis.

**Phase evolution and X-ray diffraction analysis:** The structural evolution of NbN thin films as a function of nitrogen incorporation was systematically investigated using X-ray diffraction (XRD), as summarized in **FIG. 1*b***. The diffraction patterns reveal a clear progression of phase formation that can be categorized into three distinct regimes, consistent with prior thin-film studies. In the low nitrogen regime (Region I), the films are dominated by metallic Nb, as evidenced by the prominent (110) reflection, indicating insufficient nitrogen incorporation to form a nitride phase. As the nitrogen concentration increases (Region II), new diffraction peaks corresponding to the face-centered cubic δ-NbN phase emerge, most notably the (111) and (220) reflections. This regime represents the formation of the superconducting phase, where nitrogen incorporation becomes sufficient to stabilize the Nb-N framework. At higher nitrogen concentrations (Region III), further incorporation of nitrogen leads to lattice distortion and increased structural disorder, manifested by peak broadening and the attenuation or disappearance of higher-order reflections, particularly the (220) peak.

Importantly, while both conventional sputtering and ion beam-assisted deposition exhibit this overall phase evolution, the incorporation of the energy-selective ion beam significantly alters the phase boundaries. Under identical $N_2$/Ar ratios, ion-assisted films exhibit earlier onset of δ-NbN phase formation, indicating enhanced nitridation kinetics and more efficient energy transfer during growth. This shift suggests that controlled ion energy not only accelerates nitrogen incorporation but also promotes the stabilization of the δ-phase at reduced nitrogen chemical potential.

**Superconducting transport properties:** Although X-ray diffraction confirms the formation of the superconducting δ-phase NbN, electrical transport measurements were performed to directly evaluate how ion beam assistance influences superconducting behavior. Five NbN films with comparable thickness and lateral dimensions were measured using a standard four-probe configuration, and the results are summarized in **FIG. 2**. The temperature was swept from 300 K to 2 K and back, while the resistance was continuously recorded and normalized to its value at 300 K to facilitate comparison across samples.

All films exhibit a characteristic superconducting transition: a gradual increase in resistivity upon cooling, followed by a sharp drop to an unmeasurable level, signaling the onset of superconductivity. Despite their similar thickness, the films display markedly different superconducting critical temperatures (Tc), reflecting their distinct growth conditions (**FIG. 2a**). Specifically, films corresponding to Regions I and III show significantly suppressed $T_c$, whereas those in Region II, where the δ-phase is most robust, exhibit substantially enhanced superconducting performance. These observations establish a direct correlation between phase purity and superconducting behavior. Importantly, the incorporation of the ion beam source alone is not sufficient to guarantee improved superconductivity; rather, $T_c$ is highly sensitive to the interplay between ion energy and nitrogen concentration. Under optimized conditions, the NbN film deposited with 8% $N_2$ achieves a $T_c$ of about 14.8 K, significantly exceeding the 12.4 K obtained from the conventionally sputtered counterpart. This enhancement underscores the critical role of energy-selective ion bombardment in promoting phase purity and homogeneity, and ultimately elevating superconducting performance.

To further evaluate the superconducting performance of the NbN films, we investigated their response to external magnetic fields. As a type-II superconductor, NbN exhibits strong resilience to magnetic pair breaking due to its intrinsically short coherence length(34). Temperature-dependent resistance measurements were therefore conducted under a series of applied magnetic fields, with representative results for the ion beam–assisted film deposited at 10% $N_2$ shown in **FIG. 3a**. As expected, the superconducting transition systematically shifts to lower temperatures with increasing magnetic field. Notably, the film retains a high critical temperature of about 14.8 K in zero field and maintains a $T_c$ of 12.1 K even under an applied field of 9 T, demonstrating robust superconductivity. The upper critical field, $H_{c2}$, was extracted using the Ginzburg-Landau theory, which describes the temperature dependence of the critical field as

$$H_{c2}(T) = \frac{H_{c2}(0)(1-T^2/T_c^2)}{1+T^2/T_c^2}$$

Where $H_{C2}(T)$ represents the upper critical field at each temperature and $H_{C2}(0)$ is the extrapolated zero-temperature upper critical field. Fitting the experimental data (FIG. 3b) yields an extrapolated upper critical field $H_{c2}(0) \approx 47$ T. From this value, the superconducting coherence length ξ(0) can be estimated using the relation $H_{c2}(0) = \Phi_0/(2\pi\xi(0)^2)$, where $\Phi_0$is the magnetic flux quantum. This analysis gives a coherence length ξ(0) ≈ 2.6 nm, consistent with the short coherence lengths characteristic of high-quality NbN films. Such a small coherence length reflects strong pairing robustness against magnetic perturbations and is indicative of a disordered yet electronically homogeneous superconducting state.

These properties have direct implications for device applications. The combination of high $T_c$, large upper critical field, and short coherence length is particularly advantageous for superconducting nanowire single-photon detectors (SNSPDs), where strong magnetic-field tolerance and uniform superconducting properties are critical for achieving high detection efficiency and low timing jitter. Similarly, for superconducting qubits and related quantum circuits, improved film homogeneity and reduced disorder can lead to lower microwave losses and enhanced coherence times. The ability to realize such high-performance superconducting properties in films grown at room temperature further highlights the potential of energy-selective ion beam engineering for scalable integration of NbN into next-generation quantum and cryogenic electronic platforms.

Additional superconducting parameters extracted from transport measurements are summarized in **Table 1**. Among the five samples, the ion beam-assisted film S3 exhibits the most optimized superconducting performance, characterized by a higher critical temperature (Tc about 14.8 K) and an exceptionally narrow transition width ($\Delta T$ = 0.065 K). This transition width is significantly smaller than that of the film deposited without ion beam assistance (S1) and is also narrower than values typically reported for NbN thin films grown under comparable conditions, indicating enhanced phase homogeneity and reduced electronic inhomogeneity. A clear correlation emerges between deposition conditions and superconducting properties. Films grown under non-optimized nitrogen concentrations (e.g., S2 and S5) exhibit substantially suppressed $T_c$ and broadened transitions. In contrast, samples within the optimal nitrogen range show consistently high $T_c$ and narrow $\Delta T$, underscoring the importance of balancing nitrogen incorporation with ion energy. Notably, while S1 exhibits a relatively high upper critical field, its broader transition and lower $T_c$ suggest that ion beam assistance improves not only the superconducting onset temperature but also the uniformity of the superconducting state. These results collectively demonstrate that energy-selective ion beam control enables simultaneous optimization of $T_c$, $H_{c2}$, and $\Delta T$, providing a robust pathway for enhancing superconducting film quality.

**Chemical bonding and stoichiometry analysis.** To elucidate the chemical bonding environment and elemental composition of the NbN thin films, X-ray photoelectron spectroscopy (XPS) was performed, providing insight into the degree of nitridation and the influence of ion beam assistance on local electronic structure.

For the film deposited without ion-source assistance, the Nb 3*d* spectrum exhibits the characteristic NbN doublet at binding energies of 203.93 eV and 206.53 eV, corresponding to Nb $3d_{5/2}$ and Nb $3d_{3/2}$, respectively. In addition, a secondary doublet at higher binding energies is observed and attributed to $Nb_2O_5$, which commonly forms upon air exposure of NbN surfaces(35). In contrast, the film deposited with ion beam assistance shows a systematically positive shift of approximately 0.35 eV in the Nb-N-related peaks. This shift is indicative of a modified local electronic environment, consistent with enhanced Nb-N bond strength and increased nitridation facilitated by ion-induced surface activation and densification during growth. Notably, the $Nb_2O_5$-related features remain largely unchanged, suggesting that the observed shift arises primarily from intrinsic changes in Nb–N bonding rather than variations in surface oxidation. The N 1s spectra further corroborate this interpretation. The nitride-related peak shifts from 397.65 eV (without ion assistance) to 397.48 eV (with ion assistance), corresponding to a negative shift of approximately 0.175 eV. This shift reflects an increased electron density around nitrogen, consistent with stronger Nb-N bonding and improved stoichiometric incorporation of nitrogen into the lattice. Taken together, these XPS results demonstrate that energy-selective ion beam assistance promotes a more favorable chemical bonding environment by enhancing nitridation efficiency and reducing deviations from ideal stoichiometry. Such improvements in the local bonding configuration are expected to suppress electronic inhomogeneity and scattering, thereby contributing directly to the enhanced superconducting properties observed in these films.

**Microstructural evolution and crystallization analysis.** To elucidate the role of ion-source assistance in governing film crystallization, cross-sectional transmission electron microscopy (TEM) was performed on NbN thin films. As shown in **FIG. 5a**, films deposited with ion beam assistance exhibit the onset of crystallinity much closer to the Si substrate interface compared to those grown without assistance, indicating that ion-induced energy transfer promotes early-stage nucleation and accelerates crystallization during film growth. To quantitatively assess these structural differences, grain boundary-related features were analyzed using a combination of bandpass filtering and fast Fourier transform (FFT)-based image processing (see **FIG. S1** for workflow details)(36, 37). After aligning all profiles with respect to the Si-NbN interface, the ion-assisted films display consistently higher variance intensity across the entire thickness (**FIG. 5b**), reflecting enhanced spatial ordering and reduced structural disorder. In addition, vertical column sharpness, an indicator of grain coherence and boundary definition, was extracted as a function of depth. As shown in **FIG. 5c**, ion-assisted films exhibit systematically higher sharpness values across all layers, confirming improved grain continuity and more uniform columnar growth. These microstructural improvements have direct implications for superconducting properties. In NbN, structural disorder, grain boundary scattering, and compositional inhomogeneity are known to suppress the superconducting critical temperature and broaden the transition width by introducing spatial variations in the superconducting gap(38, 39). The enhanced crystallinity and reduced disorder observed in ion-assisted films are therefore expected to minimize electronic scattering and promote a more homogeneous superconducting state. This interpretation is consistent with the experimentally observed increase in $T_c$ and the significant narrowing of the transition width ($\Delta T$). Furthermore, improved grain connectivity supports more coherent supercurrent transport across the film, which is critical for maintaining high critical fields and robust superconductivity under applied magnetic fields. Together, these results establish a clear structure–property relationship in which energy-selective ion beam assistance enhances microstructural ordering, thereby directly enabling superior superconducting performance.

**Conclusion**

We have demonstrated that energy-selective ion beam-assisted sputtering enables the room-temperature synthesis of high-quality superconducting NbN thin films with enhanced phase purity and performance. Precise control of ion energy and flux shifts the phase boundaries, promotes FCC $\delta$-NbN formation, and improves nitridation, crystallinity, and bonding. These effects collectively lead to higher critical temperatures, narrower transition widths, and robust upper critical fields, establishing a direct link between non-equilibrium growth control and superconducting functionality. More broadly, this work introduces energy-resolved ion engineering as a versatile strategy for stabilizing functional quantum materials under conditions inaccessible to conventional methods. The ability to achieve high-performance superconducting films at low temperature opens a practical pathway toward scalable integration of NbN into Complementary Metal-Oxide-Semiconductor (CMOS)-compatible quantum devices, including superconducting detectors and circuits.

**Materials and Methods**

1. Thin-film deposition. The FCC NbN thin films were deposited using a magnetron sputtering system (PVD75, *K. J. Lesker*) equipped with a load-lock chamber to enable sample transfer without breaking vacuum. A circular magnetron (TORUS TM3, *K. J. Lesker*) with a 76.2 mm diameter, 99.99% purity Nb target was employed. A broad-beam ion source (SPR-100, *Scion Plasma LLC*) was positioned 100 mm from the substrate to assist film growth. Glass and Si substrates (25.4 × 25.4 mm²) were ultrasonically cleaned in acetone (10 min), followed by rinsing in isopropanol and deionized water, and drying at 80 °C for ≥1 h prior to loading. The base pressure of the deposition chamber was maintained below 6 × 10-5 Pa. During deposition, Ar (9 sccm) and $N_2$ (3 sccm) were introduced, resulting in a working pressure of 0.25 Pa. Magnetron sputtering was operated in pulsed DC mode at 80 W with a frequency of 100 kHz and a reverse time of 1 μs. All films were deposited at room temperature unless otherwise specified. The substrate holder was rotated at 15 rpm to ensure thickness uniformity.
2. Ion beam control and diagnostics. The ion source was driven by a 13.56 MHz RF power supply (5-200 W) in combination with a DC bias (0–250 V), enabling independent tuning of ion energy and flux. Ion energy distributions and flux densities at the substrate position, which were measured using an ion analyzer (Semion 2500, *Impedans*).

3. Film thickness and growth rate calibration. Deposition rates were calibrated by depositing thick films over extended durations. Subsequent deposition times were adjusted to obtain films with thicknesses of approximately 400 nm or 1 μm for comparative studies. Film thicknesses were measured using a stylus profilometer (Dektak 150).

4. Structural characterization. X-ray diffraction (XRD) measurements were performed using a Rigaku Miniflex diffractometer with Cu Kα radiation (λ = 1.5406 Å). Cross-sectional transmission electron microscopy (TEM) was conducted using a Thermo Fisher Spectra 300 microscope. TEM specimens were prepared using a focused ion beam system (ZEISS Crossbeam 550 FIB-SEM), enabling full-thickness analysis of the NbN films.

5. Surface and chemical analysis. X-ray photoelectron spectroscopy (XPS) measurements were performed using a PHI 5000 VersaProbe II system with Al Kα radiation. High-resolution spectra were acquired with a pass energy of 23.5 eV, step size of 0.025 eV, and dwell time of 100 ms over 30 cycles. Prior to measurement, surfaces were cleaned by $Ar^+$ sputtering (500 V, 3 × 3 $mm^2$ area, 1 min). Binding energies were calibrated using the C 1s peak at 284.8 eV.

6. Superconducting transport measurements. Superconducting properties were measured using a standard four-probe configuration in a Physical Property Measurement System (PPMS, *Quantum Design*).

7. Image analysis and crystallinity quantification. To quantify structural ordering, TEM images were analyzed using custom Python scripts based on OpenCV and NumPy libraries. Images were first aligned to the Si–NbN interface for consistent depth referencing. Crystallinity was evaluated using fast Fourier transform (FFT) filtering to isolate grain boundary features. The variance of the filtered signal and vertical column sharpness were computed as a function of depth to assess spatial evolution of crystallinity across the film thickness.

**Acknowledgments**

The authors are thankful for the support from the following programs: NSF award #2243110, the MEDC TSGTD and ADVANCE programs, and AAP-PIRA program. This work done by R.J.C. is supported by the U.S. Department of Energy, Office of Science, National Quantum Information Science Research Centers, Co-design Center for Quantum Advantage ($C^2QA$) under contract number DE-SC0012704. M.X. and W.X. were supported by U.S.DOE-BES under Contract DE-SC0023648.

## Figures and Tables

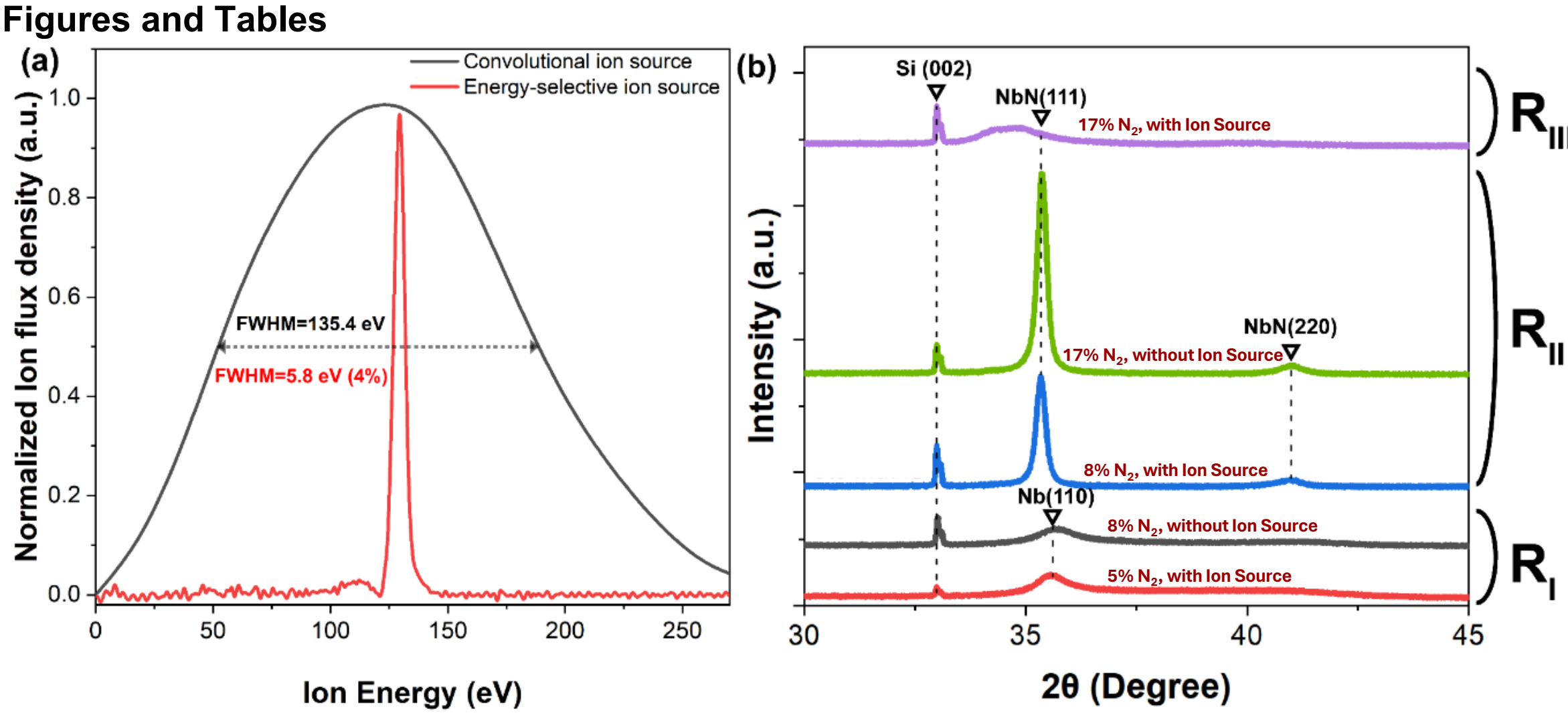

**Figure 1.** (a) Ion energy distributions measured for a conventional anode-layer ion source and the energy-selective ion beam source, both operated at identical nominal ion energies. The energy-selective source exhibits a significantly narrower distribution. (b) X-ray diffraction (XRD) patterns of 250 nm thick NbN thin films deposited on Si substrates at varying nitrogen concentrations, with and without ion beam assistance. Regions I-III (RI-RIII) denote distinct structural regimes corresponding to metallic Nb (RI), formation of face-centered cubic δ-NbN (RII), and nitrogen-rich, structurally disordered phases (RIII).

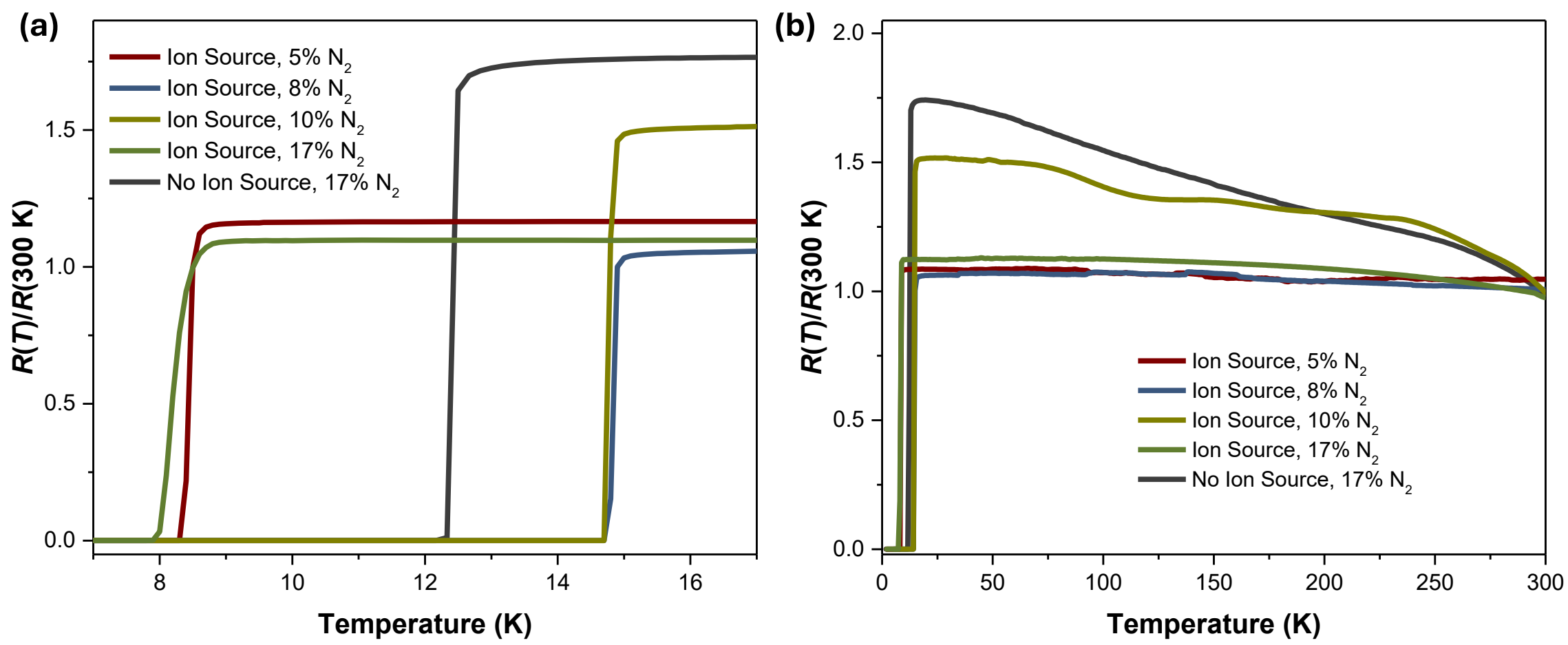

**Figure 2.** (a) Temperature dependence of the normalized resistance (R/R300K) for 250 nm thick Nb-N thin films deposited under varying nitrogen concentrations, with and without ion beam assistance, shown in the vicinity of the superconducting transition. (b) Full temperature-dependent normalized resistance curves for the same films measured over the range 2-300 K, illustrating the complete evolution from normal-state transport to the superconducting state.

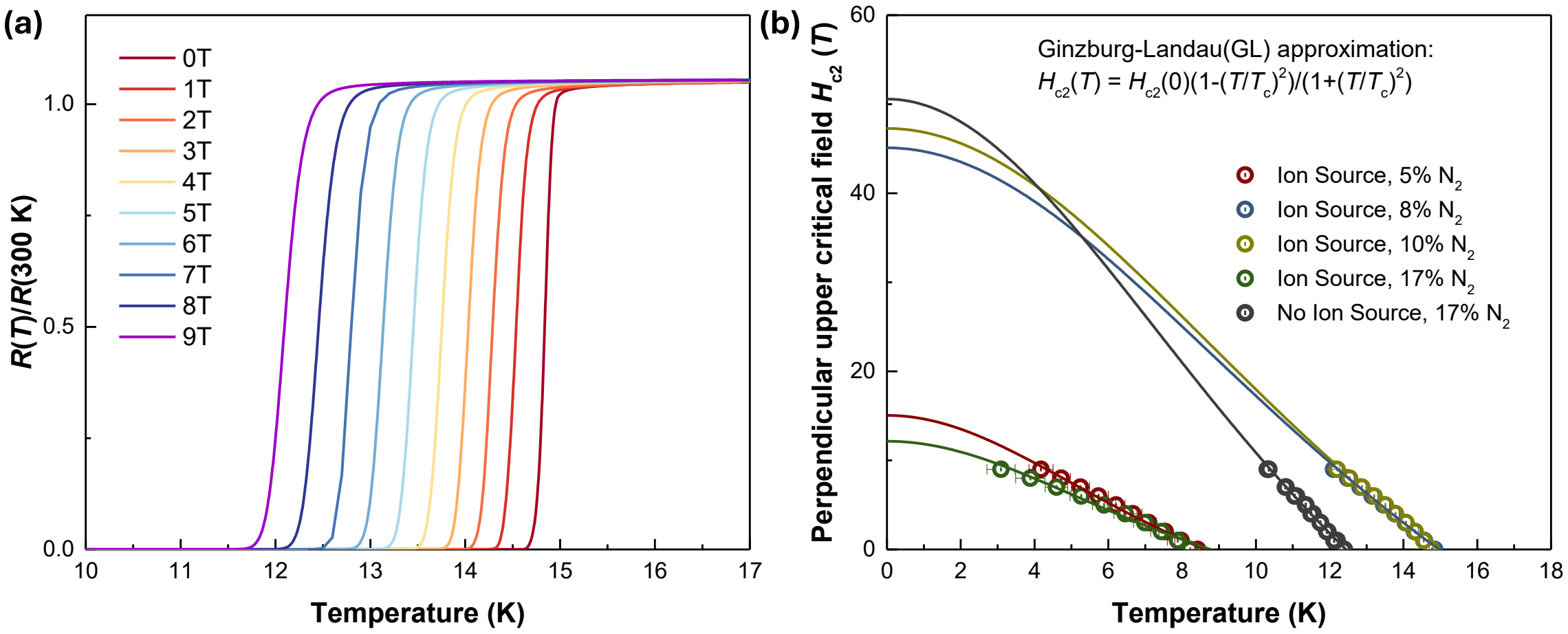

**Figure 3.** (a) Temperature dependence of the normalized resistance ($R/R_{300}$K) for an NbN thin film deposited at 10% nitrogen concentration with ion beam assistance, measured under varying applied magnetic fields, highlighting the field-induced suppression of the superconducting transition. (b) Upper critical field ($H_{c2}$) as a function of temperature for Nb-N thin films extracted from transport measurements under applied magnetic fields. The solid line represents the fit based on the Ginzburg-Landau theory.

**Table 1.** Deposition parameters and superconducting properties of NbN thin films. Listed are the ion source conditions, nitrogen concentration, superconducting critical temperature ($T_c$), upper critical field at zero temperature [Hc2(0)], and superconducting transition width ($\Delta T$).

| Sample | Ion Source | $T_c$ | $N_2$ Concentration | $H_{C2}(0)$ | $\Delta T$ |
|---|---|---|---|---|---|
| S1 | No | 12.44 K | 17% | ~51 T | 0.092 K |
| S2 | 20W RF power, 60V DC voltage, 200mA | 8.22 K | 17% | ~12 T | 0.418 K |
| S3 | 20W RF power, 60V DC voltage, 200mA | 14.79 K | 10% | ~47 T | 0.065 K |
| S4 | 20W RF power, 60V DC voltage, 200mA | 14.84 K | 8% | ~45 T | 0.091 K |
| S5 | 20W RF power, 60V DC voltage, 200mA | 8.44 K | 5% | ~15 T | 0.133 K |

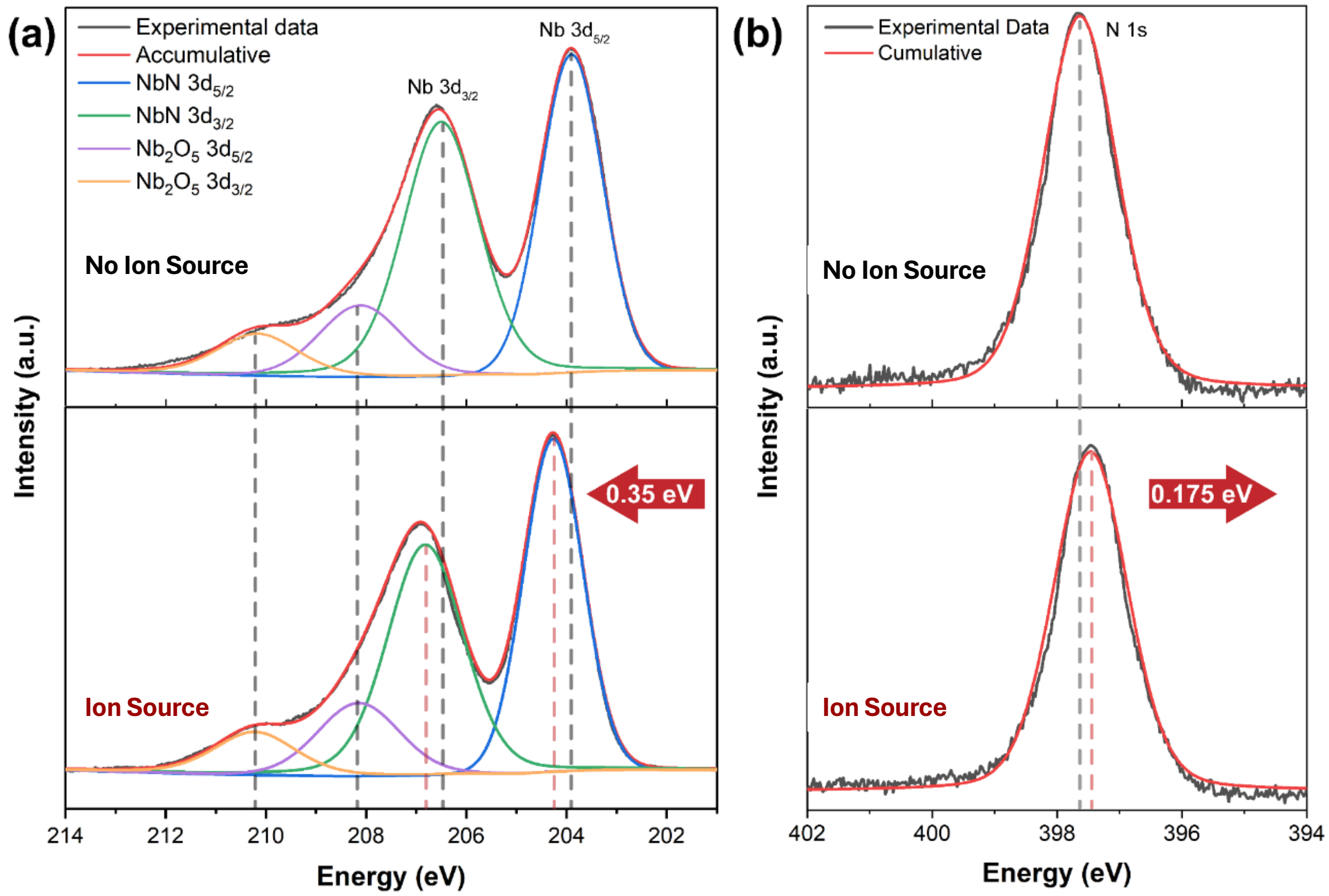

**Figure 4.** High-resolution X-ray photoelectron spectroscopy (XPS) spectra of NbN thin films deposited with and without ion source (IS) assistance. (a) Nb 3d and (b) N 1s regions. The experimental data (black) are deconvoluted into individual spin-orbit and chemical components (colored curves), with the overall fitted envelope shown in red. The spectra reveal shifts in binding energy associated with ion beam–induced modifications of the Nb-N bonding environment.

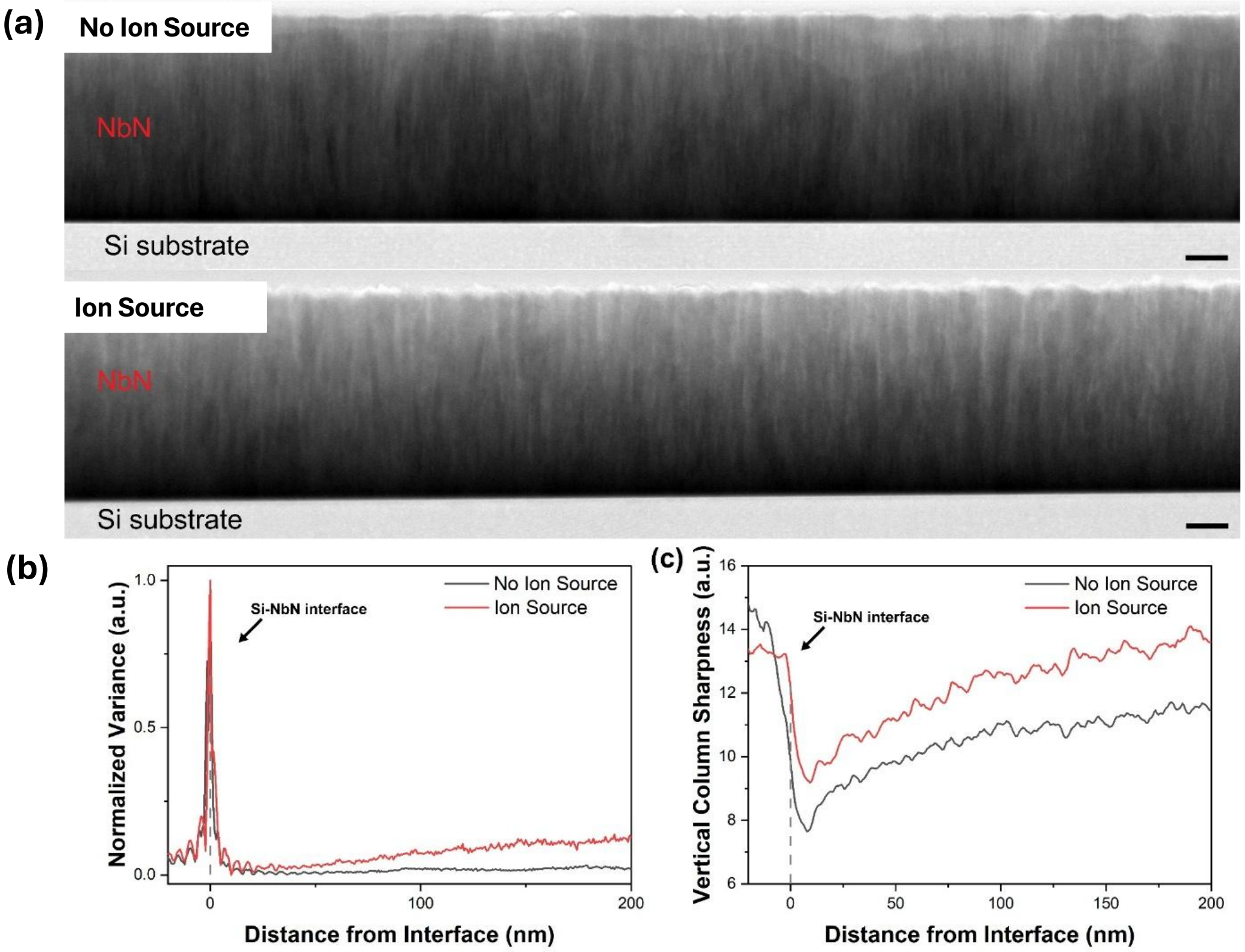

**Figure 5.** Transmission electron microscopy (TEM) analysis of crystallization in NbN thin films. (a) Cross-sectional TEM images of films deposited with and without ion source assistance, showing earlier onset of crystallinity and improved columnar structure in the ion-assisted film. (b) Depth-dependent profiles of normalized variance extracted from grain boundary-related features, aligned with respect to the Si-NbN interface, indicating enhanced spatial ordering in the ion-assisted sample. (c) Corresponding profiles of vertical column sharpness as a function of depth, demonstrating improved grain definition and continuity. Scale bar: 50 nm.

## Supporting Information for

# Low-Temperature Stabilization of δ-NbN Superconducting Thin Films through Energy-Selective Ion Beam Sputtering

Yifan Liu[1,2], Junwoo Lee[3], Mingyu Xu[4], Weiwei Xie[4*], Robert J. Cava[5*], Qi Hua Fan[1,3*]

1. Department of Electrical and Computer Engineering, Michigan State University, East Lansing, MI 48824, USA

2. Institute for Quantitative Health Science and Engineering, Michigan State University, East Lansing, MI 48824, USA

3. Department of Chemical Engineering and Materials Science, Michigan State University, East Lansing, MI 48824, USA

4. Department of Chemistry, Michigan State University, East Lansing, MI 48824, USA

5. Department of Chemistry, Princeton University, Princeton, NJ 08540, USA

* Robert J. Cava, Qi Hua Fan, Weiwei Xie

**Email:** rcava@princeton.edu; qfan@egr.msu.edu; xieweiwe@msu.edu

## Supporting Information Text

**Grain boundary feature extraction and crystallinity quantification.** The workflow for extracting grain boundary–related features from cross-sectional TEM images is illustrated in **FIG. S1**. The original images were first cropped and spatially aligned such that the Si-NbN interface was consistently registered across all samples. Fast Fourier transform (FFT) was then applied to obtain the corresponding power spectra. A bandpass filter (0.045-0.281 nm-1) was used to selectively remove low-frequency background contributions and high-frequency noise, thereby isolating structural features associated with grain boundaries and interfaces. The filtered spectra were subsequently transformed back into real space via inverse FFT, yielding images in which grain boundary networks and the Si-NbN interface are emphasized. Finally, the variance of the filtered signal was computed as a function of depth to quantitatively assess the spatial evolution of crystallinity throughout the film thickness.

**High-angle powder X-ray diffraction pattern:** To further confirm the phase purity of the optimized NbN films, high-angle X-ray diffraction patterns were collected and are presented in **Fig. S2**. In addition to the strong Si substrate reflections, well-defined FCC NbN (420) and (422) reflections are observed for both the conventionally sputtered and ion beam–assisted films. The ion beam–assisted film deposited with 8% $N_2$ exhibits sharp and intense NbN reflections despite the substantially lower nitrogen concentration, confirming that energy-selective ion beam assistance effectively stabilizes the δ-NbN phase. These results further support the phase evolution discussed in the main text and demonstrate that ion beam assistance promotes the formation of highly crystalline superconducting NbN at reduced nitrogen concentrations.

## Figures

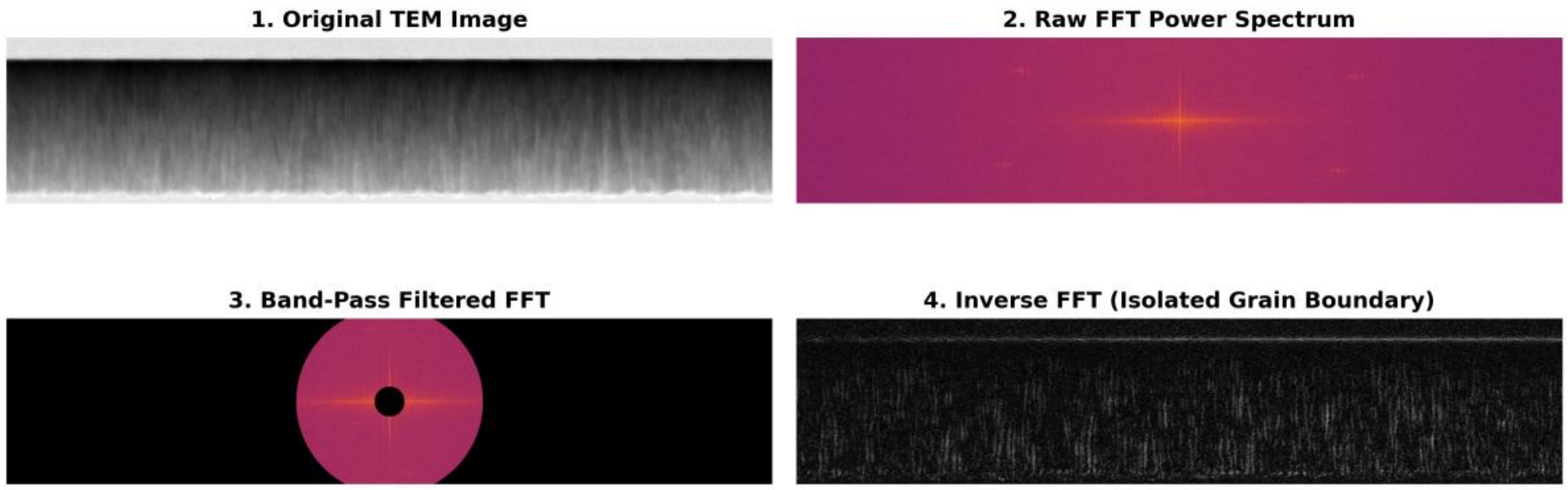

**Fig. S1.** Workflow for extracting grain boundary–related features from cross-sectional TEM images using bandpass filtering combined with fast Fourier transform (FFT) analysis. The procedure isolates structurally relevant frequency components, enabling quantitative comparison of crystallinity and spatial ordering across different samples.

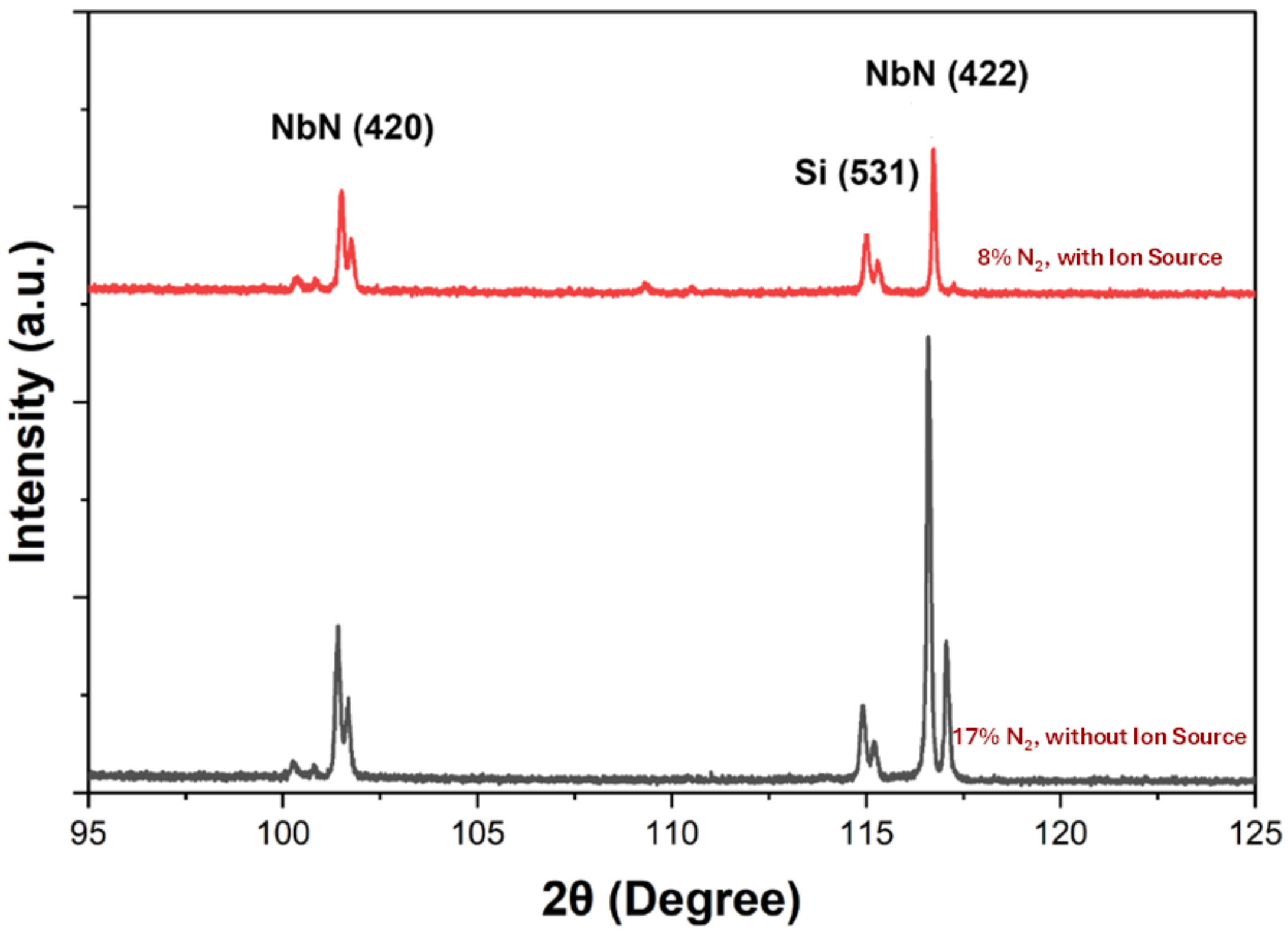

**Fig. S2.** High-angle X-ray diffraction (XRD) patterns of representative NbN thin films deposited under optimized conditions with and without ion source (IS) assistance. The ion beam–assisted film (8% $N_2$ with IS) exhibits well-defined FCC NbN (420) and (422) reflections comparable to those of the conventionally sputtered film (17% $N_2$ without IS), despite being deposited at a substantially lower nitrogen concentration. The Si (531) reflection originates from the Si substrate. Peaks around 110 degrees may be associated with $NbO_xN_y$ minor phases. This is related to the high-energy ion bombardment induced by the ion source. This bombardment promotes the crystallization of the sputtered Nb atoms in the oxygen-containing interfacial layer. Finally, we note the presence of the additional reflection near the (422) peak for the non-IS 17% sample. Its position is close to expectations for the alpha 1-2 splitting, but its intensity and peak shape are not. Its presence may be due to a subtle difference in symmetry for an $NbN_x$ phase deposited under these conditions, leading to its lower Tc, but higher precision data than is available to us may be needed to resolve this issue.